\documentclass[a4paper,11pt]{article}
\usepackage{jinstpub} 
\usepackage{lineno}

\title{\boldmath Recent results on the low-pressure GEM-based TPC at an Accelerator Mass Spectrometer}

\author[a,b]{A. Bondar,}
\author[a,b]{V. Parkhomchuk,}
\author[a,b]{A. Petrozhitsky,}
\author[a,b]{T. Shakirova,}
\author[a,b]{A. Sokolov}

\affiliation[a]{Budker Institute of Nuclear Physics,\\11, Acad. Lavrentieva Pr., Novosibirsk, 630090, Russia}
\affiliation[b]{Novosibirsk State University,\\1, Pirogova str, Novosibirsk, 630090, Russia}

\emailAdd{T.M.Shakirova@inp.nsk.su}

\abstract{
The Accelerator Mass Spectrometry technique makes it possible to measure rare long-lived isotopes such as $^{10}$Be, $^{14}$C, $^{26}$Al, $^{36}$Cl, $^{41}$Ca and $^{129}$I. The content of these isotopes can be at the level of 10$^{-15}$ of the total element content. The Accelerator Mass Spectrometer developed by Budker Institute of Nuclear Physics (BINP AMS) successfully measures the concentration of $^{14}$C relative $^{12}$C. However, there is a problem of separating the $^{10}$B isobaric background from $^{10}$Be. Beryllium-10 is used to date geological objects on a time scale from 1 thousand years to 10 million years.
	
To solve this problem we have proposed a new technique for ion identification based on measuring both ion track ranges and ion energies in a low-pressure Time-Projection Chamber (TPC) with Gas Electron Multiplier (GEM) readout. We have developed the TPC with a dedicated thin silicon nitride window for an efficient passage of ions. To begin with, the characteristic of the low-pressure TPC were studied in isobutane at a pressure of 50 torr using alpha particle sources. 
	
In this work, we set up the low-pressure TPC on BINP AMS facility and successfully measured track ranges and energies of ions from samples containing $^{14}$C. At the next stage, we are going to carry out measurements with samples containing $^{10}$Be. However, using the obtained results and SRIM simulation we have already shown that the isobaric boron and beryllium ions can be separated by more than 5 sigma. This technique is proposed to be applied in AMS for dating geological objects, namely for geochronology of Cenozoic era.}


\begin{document}
\maketitle
\flushbottom

\section{Introduction}
\label{sec:intro}
Accelerator mass spectrometry (AMS) allows to measure the content of rare long-lived cosmogenic and antropogenic isotopes in a sample, such as $^{10}$Be, $^{14}$C, $^{26}$Al, $^{36}$Cl, $^{129}$I and others. These isotopes are formed as a result of the interaction of cosmic radiation with the atmosphere or the Earth surface. They can also be formed at nuclear power plants or during nuclear explosions~\cite{cosmog_isotopes}. AMS makes it possible to measure the concentration of these isotopes at their fraction of 10$^{-15}$ of the total isotope content. This technique combines the high sensitivity of mass spectrometric methods with a high level of molecular and ion background suppression.

The best-known example of the widespread use of AMS is radiocarbon dating, in which $^{14}$C is compared with its stable isotopes $^{12}$C and $^{13}$C. With radiocarbon dating, it is possible to restore the age of a sample in the time interval from 300 years to 50 thousand years, using a small amount of substance (<1 mg). Today, the AMS technique is used not only for radiocarbon dating, but has also become widespread in other fields, such as archaeology, pharmacology, geology, ecology, forensic science, etc.

There are over 140 AMS complexes operating in the world, one of which was developed at the Budker Institute of Nuclear Physics (BINP)~\cite{BINP_AMS}. The main task in creating AMS complex is to isolate the cosmogenic isotope of interest from the accompanying background, consisting of isobars and molecules with masses close to the isotope being measured. Isobars are atoms of different chemical elements that have the same mass number. Therefore AMS complexes combine a number of methods in order to suppress this background.

Successful measurements of $^{14}$C concentration are being carried out at the BINP AMS facility using a time-of-flight system~\cite{BINP_TOF}. Additionally, the design of the BINP AMS allows for measurements of other rare cosmogenic isotopes. However, there is a problem in separating isobaric ions, which form negative ions in the ion source, such as $^{10}$B in the case of cosmogenic $^{10}$Be. This isobaric ion cannot be separated using existing methods at the AMS facility. Beryllium-10 is particularly significant in environmental sciences (archaeology, glaciology, oceanography) and geology for dating purposes within the time range of 1 thousand to 10 million years. To solve this problem, we propose a new technique for ion identification based on measuring ion track ranges in a low-pressure time projection chamber (TPC).

\section{New concept of ion identification at AMS}
\label{sec::SRIM}
Semiconductor detectors, time-of-flight systems and ionization chambers are conventionally used at AMS complexes. The semiconductor detector cannot be used to separate isobaric ions, since it measures only the energy of ions. In addition, they are susceptible to radiation damage. Time-of-flight systems can only provide the separation of isotopes, since ions with different masses have different velocities~\cite{AMS_appl}. Ionization chambers with a segmented anode and a transverse electric field are used to separate isobaric ions. The disadvantage of this type of detector is a poor signal/noise ratio, since the chamber works in the ionization mode, which makes high requirements for electronics.

A new concept of identifying isobaric ions at AMS is to separate them by their track ranges in a low-pressure TPC with the charge signal amplification using a thin gas electron multiplier (GEM). This is possible, since ions with different atomic numbers have different energy losses in matter. The possibility of separating $^{10}$Be and $^{10}$B isobars was demonstrated using the SRIM software package~\cite{srim}. The results of track range simulations in isobutane at a pressure of 50 Torr, taking into account the entrance membrane made of Si$_{3}$N$_{4}$ with a thickness of 200 nm show that ion track ranges differ by approximately 12 mm. Therefore, the low-pressure TPC with a spatial resolution of the order of 2–3mm will be sufficient to separate such ions.

To determine the spatial resolution of the low-pressure TPC, the detector was calibrated using the alpha particles source from $^{226}$Ra. This source emits monochromatic alpha particles at five different energies: 4.8 MeV, 5.3 MeV, 5.5 MeV, 6 MeV and 7 MeV. The track ranges in isobutane from alpha particles were also simulated under the same conditions. Alpha particles with energies of 5.3 and 5.5 MeV have the same track range difference as boron and beryllium, about 12 mm. If we can separate such tracks using the low-pressure TPC, then we will be able to identify tracks from $^{10}$Be and $^{10}$B ions at the BINP AMS.

To test the feasibility of track range separation for isobaric boron and beryllium ions, a low-pressure TPC was developed. The detector's operating principle and previous results are detailed in~\cite{LpTPC_1} and~\cite{LpTPC_2}.

\section{Experimental results with alpha particle source}
\label{sec::alpha_measur}

\begin{figure}[htbp]
	\centering
	\includegraphics[width=.45\textwidth]{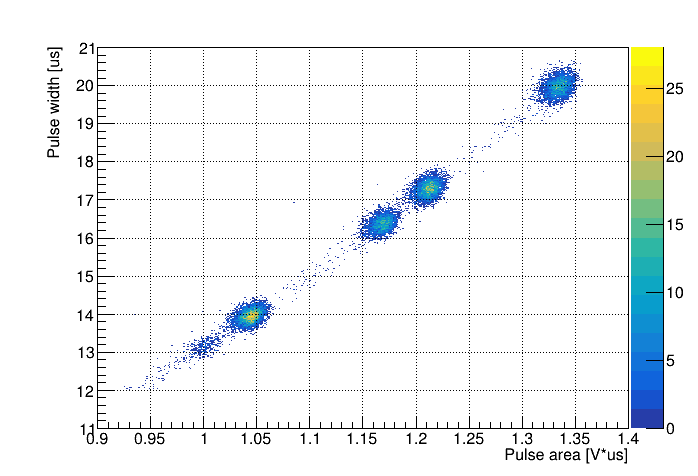}
	\qquad
	\caption{2D plot of pulse width versus pulse area for alpha particles from $^{226}$Ra source, measured in low-pressure TPC in isobutane at 50 torr and room temperature using GEM amplification with gain of 320.}
	\label{fig:2D_plots_alphas}
\end{figure}

The alpha particle spectrum from the $^{226}$Ra radioactive source was measured in the low-pressure TPC in isobutane at 50 torr pressure, with an effective gain of the GEM set to 320. Since alpha particles at a nominal pressure of 50 torr have long track ranges, a large-size low-pressure TPC was used to determine the detector's resolution with $^{226}$Ra. This detector has a length of 30 cm and a diameter of 10 cm. Figure~\ref{fig:2D_plots_alphas} shows a two-dimensional distribution of pulse width and area, where four clusters corresponding to alpha particles with energies of 4.8, 5.3, 5.5, and 6 MeV can be observed. The alpha particle with an energy of 7.7 MeV has a longer track range than the detector's length. As depicted in figure~\ref{fig:2D_plots_alphas}, these clusters are well-separated both by ion track range and energy.
 
The clusters in the center belong to alpha particles with energies of 5.3 and 5.5 MeV and have the same difference in track range - 12 mm, as previously mentioned. These alpha particle lines can be separated based on track range at the 5 sigma level. Using these experimental results and SRIM simulation, it has been shown that $^{10}$B and $^{10}$Be isobars can also be effectively separated at the 5 sigma level in the AMS complex using the low-pressure TPC.

\section{Experimental results with ions from AMS}
\label{sec:AMS_measur}
After successful testing of the low-pressure TPC with the alpha particle source, the detector was installed at the BINP AMS facility after the time-of-flight system. Therefore, debugging our detector on the AMS will not affect routine measurements. We carried out test measurements with two types of samples: standard and background. The standard sample is an Australian National University (ANU) standard made from sucrose and is one of the certified radiocarbon standards. The precise content of $^{14}$C is known in such samples. The background sample was made from fine-grained dense graphite, where, with proper sample preparation, the concentration of $^{14}$C is very low. During routine sample analysis, all results are corrected for background and normalized to the standard samples.

\begin{figure}[htbp]
	\centering
	\begin{minipage}{0.45\textwidth}
		\centering
		\includegraphics[width=0.9\textwidth]{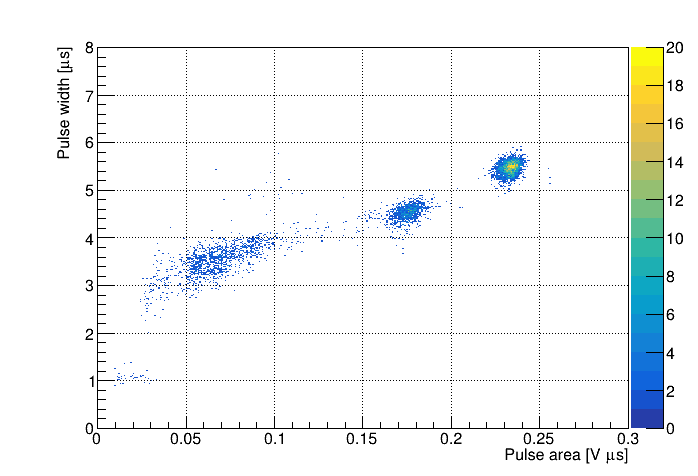}
		\caption{2D plot of pulse width versus pulse area for ions from BINP AMS measured in low-pressure TPC in isobutane at 50 torr. Ions were extracted from reference sample (Australian National Standard) where the exact content of radiocarbon is known.}
		\label{fig:2D_plots_ANU}
	\end{minipage}\hfill
	\begin{minipage}{0.45\textwidth}
		\centering
		\includegraphics[width=0.9\textwidth]{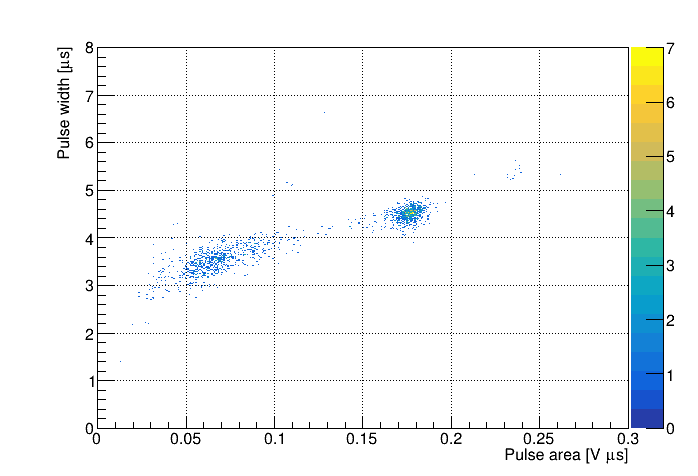}
		\caption{2D plot of pulse width versus pulse area for ions from BINP AMS measured in low-pressure TPC in isobutane at 50 torr. Ions were extracted from background sample (fine-grained dense graphite). There are not $^{14}$C.}
		\label{fig:2D_plots_blank}
	\end{minipage}
\end{figure}

Two-dimensional distributions of pulse width and areas of ions from BINP AMS are presented in figures~\ref{fig:2D_plots_ANU} and~\ref{fig:2D_plots_blank}, respectively, measured in the low-pressure TPC. In figure~\ref{fig:2D_plots_ANU}, we observe three clusters from the standard sample, while in figure~\ref{fig:2D_plots_blank}, from the blank sample, we observe two clusters. Thus, it can be concluded that the peak of ions in the 5.5 $\mu$s region belongs to $^{14}$C ions. The other two clusters represent background, which the time-of-flight detector installed at the BINP AMS successfully suppresses.

\begin{figure}[htbp]
	\centering
	\includegraphics[width=.45\textwidth]{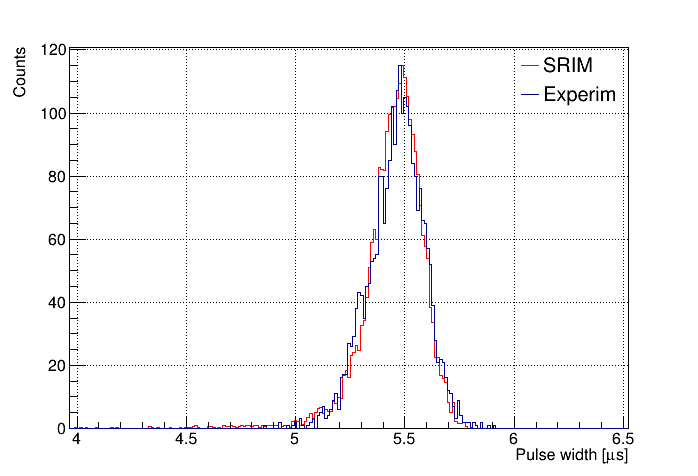}
	\qquad
	\caption{Distribution of pulse width from $^{14}$C ions. Low-pressure TPC measurements are fitted with simulation using SRIM.}
	\label{fig:exp_vs_simul}
\end{figure}

A distribution of pulse width from $^{14}$C ions was obtained, as shown in figure~\ref{fig:exp_vs_simul}. Measurements in the low-pressure TPC were fitted using SRIM. The drift velocity of electrons from the experiment is 0.53 cm/$\mu$s. Using Garfield++\cite{garfield} and Magboltz\cite{magboltz}, the drift velocity of electrons in isobutane at a pressure of 50 torr and an electric field intensity of 13.5 V/cm was calculated. The drift velocity of electrons in isobutane was found to be 0.58 cm/$\mu$s. In Figure~\ref{fig:exp_vs_simul}, the mean values of the experimental data and simulation are manually aligned for comparison. The mean of experimental data differ from the simulation mean by 9\%, while the distribution width is the same. Thus, the experiment and simulation show reasonable agreement.

The deviation of the simulation from the experiment may be caused by the straggling of energy losses in the nitrocellulose films installed in the time-of-flight detector of the BINP AMS. In addition the inaccuracy in the absolute ion energy at the accelerator output may lead to the observed differences in ion ranges between simulation and experiment.

\section{Conclusion}
\label{sec:conclusion}

We have developed and successfully tested the low-pressure TPC with GEM readout for AMS. A new ion identification technique based on measuring both ion track ranges and ion energies in low-pressure TPC has been developed. Using these results and SRIM simulations, it was shown that isobaric boron and beryllium ions can be effectively separated at BINP AMS at a level of 5 sigma. It provides efficient dating up to 10 million years (for geochronology of Cenozoic Era). 

Low-pressure TPC was installed at BINP AMS. Measurements were successfully carried out with samples containing radiocarbon. After the modernization of BINP AMS we plan to conduct experiments with samples containing beryllium.

\acknowledgments

This work was supported in part by Russian Science Foundation (project no. 23-22-00359).



\begin{thebibliography}{99}

\bibitem{cosmog_isotopes}
G. Vagner, \emph{Age Determination of Young Rocks and Artifacts: Physical and Chemical Clocks an Quaternary Geology and Archeology}, \emph{Springer-Verlag Berlin Heidelberg} (1998).	
	
\bibitem{BINP_AMS}
V. V. Parkhomchuk, S. A. Rastigeev, \emph{Accelerator mass spectrometer of the center for collective use of the Siberian Branch of the Russian Academy of Sciences}, \emph{Journal of Surface Investigation. X-ray, Synchrotron and Neutron Techniques} {\bf 5} (2011) 1068-1072.
	
	
\bibitem{BINP_TOF}
N. I. Alinovskii et al., \emph{A time-of-flight detector of low-energy ions for an accelerating mass-spectrometer}, \emph{General Experimental Techniques}, {\bf 52} (2009) 234-237.
	
\bibitem{AMS_appl}
L. K. Fifield, \emph{Accelerator mass spectrometry and its applications}, \emph{Reports on Progress in Physics}, {\bf 62} (1999) 1223.
	
\bibitem{srim}
http://www.srim.org

\bibitem{LpTPC_1}
A. Bondar et al., \emph{Low-pressure TPC with THGEM readout for ion identification in Accelerator Mass Spectrometry}, \emph{Nucl. Instrum. Methods A}, {\bf 958} (2020) 162780.

\bibitem{LpTPC_2}
A. Bondar et al., \emph{New technique of ion identification in Accelerator Mass Spectrometry using low-pressure TPC with GEM readout}, \emph{JINST} {\bf18} (2023) C05015.
	
\bibitem{Si3N4_membrane}
M. Dobeli et al., \emph{Gas ionization chambers with silicon nitride windows for the detection and identification of low energy ions}, \emph{Nucl. Instrum. Methods B} {\bf 219-220} (2004) 415-419.
	
\bibitem{garfield}
https://garfieldpp.web.cern.ch/garfieldpp/
	
\bibitem{magboltz}
https://magboltz.web.cern.ch/magboltz/

\end{thebibliography}


\end{document}